# Survey on Energy-Efficient Techniques for Wireless Sensor Networks

*Anum Masood[1]*

[1] Computer Science Department, COMSATS University Islamabad, Islamabad, Pakistan

*Abstract*— **The concept of energy efficient computing is not new but recently the focus of the industries related to technology has been shifted towards energy utilization techniques with minimum energy loss. Computer Networks also needed to be energy efficient. Energy Consumption is also an important challenge in Wireless Sensor network. Energy efficiency can be achieved by the help of clustering algorithms, energy efficient routing methods, improved data aggregation schemes and energy efficient MAC protocols for WSN.**

1. INTRODUCTION

In wireless technology, Wireless Sensor Networks have appeared as an important new field. Wireless Sensor Networks are used in applications (Abbasi & Younis, 2007) such as surveillance, habitat supervision, intruder detection, health care, disaster-warning systems, defence systems, target tracking and security (Akyildiz, Su, Sankarasubramaniam, & Cayirci, 2002).

Researches in field of low-power radio frequency high-speed broadband wireless network technologies have given rise to the development of Wireless Sensor Networks. Wireless networks have been frequently used when there is an issue of non-access to the conducted environment (Bandyopadhyay & Coyle, 2003). In such inaccessible environment the user is usually unable to collect data; therefore Wireless Sensor Networks (WSNs) are used instead for sensing such regions. There is a limited amount of energy available hence there is need for minimizing the consumption of energy which will increase the lifetime of the battery.

In WSN, a sensor node is composed of sensing unit, processing unit, power unit and communication unit. Sensing Unit senses response to any change in physical state such as temperature, pressure, speed etc (Hill, 2003). Processing Unit collects the information about changes and process signals which were sent from sensor units. Processing Unit transmits these signals to the network (Vieira, Coelho, Silva, & Mata, 2003). Power Unit deals with the residual energy of the nodes (Feng, Koushanfar, & Potkonjak, 2002). Communication unit is supported by wireless communication channel which provides transmission medium to transfer signals from sensor unit to the network or any other output device (Heinzelman, Kulik, & Balakrishnan, 1999). Early work on energy efficient techniques introduced two methods, DPM and DVS. Dynamic Power Management (DPM) states that devices which were not being used should be temporarily shut down and re-activated when needed (Kimura, Jolly, & Latifi, 2006). But its limitations were that it requires support from the operating system and stochastic analysis was also required to predict future events as to when the devices will be required.

Second method was Dynamic Voltage Scheduling (DVS) in which power was varied to allow for a non-deterministic workload on the WSN (Yu, Moh, Lee, & Youn, 2002). By changing the voltage and the frequency, the total power consumption was effectively reduced. Predicting future workload is an important aspect of DVS. The current workload as well as the prediction about the future workloads play vital role in the decisions to spread workload. The efficiency of DVS is immensely dependent on the accuracy of the future workload of nodes prediction. Therefore the performance of DVS depends on the algorithm which is used for estimation of future workloads. For embedded systems such as wireless sensor networks, energy conservation is crucial.

In a WSN, there are various inexpensive nodes but the interaction between these nodes is restricted to limited area. The sensing capability of these nodes is also limited. Source of energy for these nodes is mostly battery (Emanuele Lattanzi, 2007). Battery's lifetime defines the lifetime of the WSN, (Kyung Tae Kim, 2009) thus the energy consumption is an important challenge (Mhatre, Rosenberg, Kofman, Mazumdar, & Shroff, 2005). The network has to be energy efficient so that it ensures prolonged lifetime of a WSN (Trigoni, Yao, Demers, Gehrke, & Rajaramany, 2004).

WSN challenges are related to energy efficient mechanisms, deployment strategies, data aggregation, coverage, tracking, security issues etc. Many applications depend on WSN for information retrieval. The self-organizing WSN have provided sensor nodes with the ability to learn by the help of adaptive algorithms while the use of adaptive power control in the IP networks uses sleep-mode, robustness, Quality of Service (QoS) for guaranteed delivery, fault-tolerance, security measures and reactive routing protocols (García-Hernández, Ibargüengoytia-González, García-Hernández, & Pérez-Díaz, 2007).

WSN may be homogeneous or heterogeneous in nature. In homogeneous WSN all the sensor nodes are same while in heterogeneous WSN different type of sensor nodes are used (Corchado, J.Bajo, Tapia, & A.Abraham, 2010). In order to fulfil the needs of various applications of sensor networks, researches on heterogeneous WSNs have been done (Duarte-Melo & Liu, 2002). Heterogeneous WSN have complex nature of energy configuration therefore reduction of energy dissipation is a challenge (Dietrich & Dressler, 2009).

WSNs are used in pervasive computing due to their flexibility, minimum energy loss and low cost. In WSN, data transmission consumes more energy as compared to data processing (Raghunathan, Schurghers, Park, & Srivastava, 2002). Energy required to transmit a data packet is approximately same as the energy required to process thousand functions in sensor node (Pottie & Kaiser, 2000). The type of sensor node used specifies the energy consumption of the sensing part of WSN. The communication among the nodes consumes most of the energy of WSN.

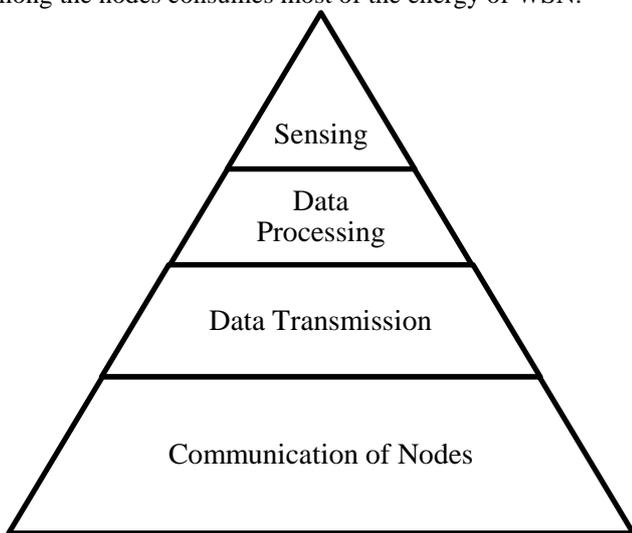

Figure 1: Energy Consumption in various phases of WSN

The energy used for sensing is negligible as compared to processing and node communication but it is greater than the energy required for the data transmission. Energy efficient techniques mainly focus on the protocol used in WSN and the sensing part of WSN. For better results and prolonging of life-span of WSN different energy efficient techniques maybe used in combination (Anastasi, Conti, Francesco, & Passarella, 2013).

The energy efficient methods have been widely used in WSNs. In Section 2, energy efficient clustering methods which enhance the lifetime of WSN are discussed. The energy efficient methods for routing in WSNs are reviewed in Section 3. Energy Efficient MAC protocols for wireless sensor networks (WSNs) are described in Section 4. Nodes have limited power hence at the MAC level energy efficiency is of immense importance. Data Aggregation is a technique for reduction of communication overhead. These energy efficient data aggregation methods are discussed in Section 5.

## 2. ENERGY- EFFICIENT CLUSTERING METHODS IN WSNs

As the durability of the wireless sensor network depends on the energy source's lifetime therefore researchers have introduced many methods for better energy efficiency. Energy efficiency is a fundamental aspect in the WSN applications as they require a prolonged lifetime with minimum energy losses.

### 2.1. Clustering Method

Clustering is a vital energy-saving technique in WSN, and the performance of WSN depends on energy-efficient clustering method. There have been numerous algorithms introduced for clustering in WSN.

The use of cluster has reduced interaction distance for the sensor nodes in WSN. There are some base-stations (BS) in the WSN which enable distant interactions between the nodes, and a cluster head to supervise each cluster. A cluster based wireless sensor network is shown in figure 2. A cluster-based protocol is used to divide the whole WSN into different clusters.

Clustering is an effective technique for topology control to solve energy efficiency issues. To achieve the goal of less energy consumption and more life-span, clustering methods have been proposed for WSNs. Clustering is performed to optimize the coverage of network, energy efficiency, distance between any two node clusters, and interaction between nodes(Olutayo Boyinbode, 2010) (Jiang, Yuan, & Zhao, 2009).

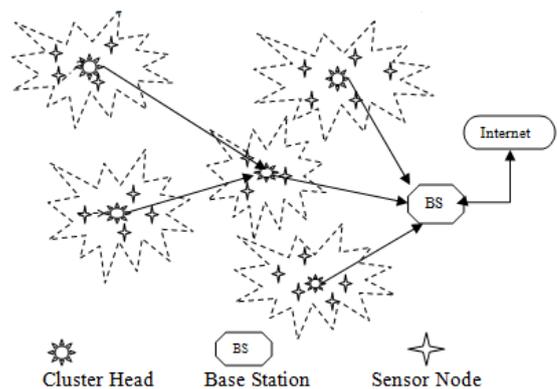

Figure 2: Wireless Sensor Network Cluster-Based

Mostly clustering algorithms are classified into two broad categories i.e. distributed clustering algorithms and centralized clustering algorithms as shown in figure 3.

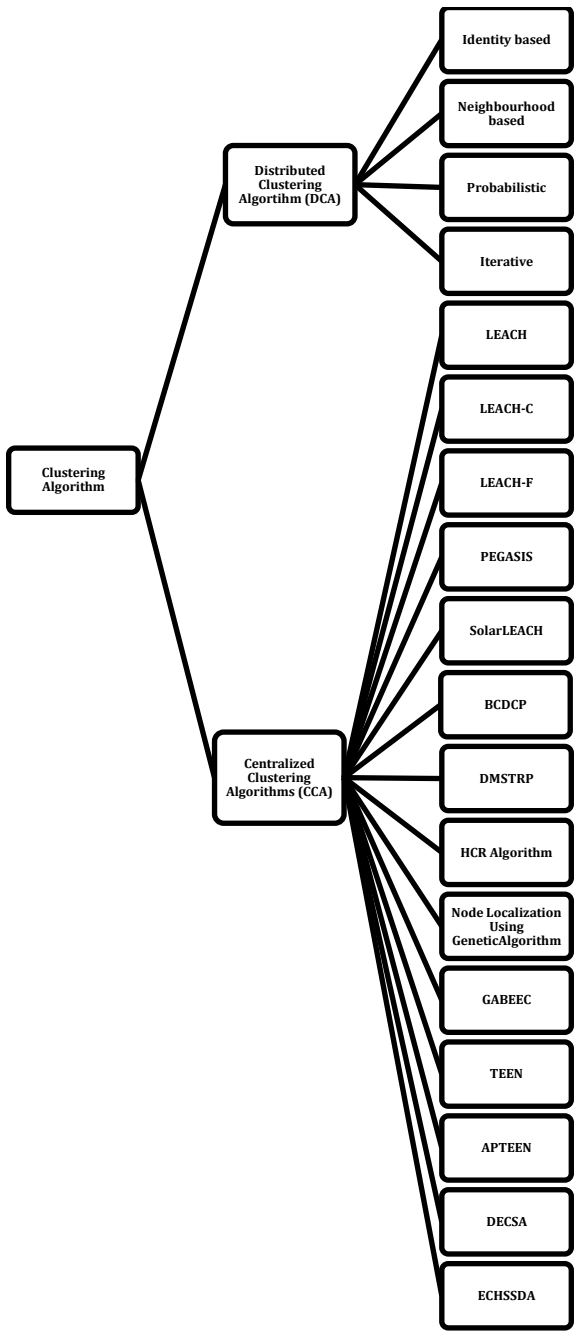

Figure 3: Classification of Clustering Algorithms in WSN

## 2.2. Distributing Clustering Algorithm:

Distributed Clustering Algorithms are generally of four types. These are given below:

### 2.2.1. Identity Based Clustering Algorithm:

Identity based clustering takes unique node identification as basic factor to select the cluster heads. Linked Cluster Algorithm (LCA) is an example of identity based clustering algorithm (Kumarawadu, Dechene, Luccini, & Sauer, 2008). Improved form of Linked Cluster Algorithm is LCA2 which ensures that there will be no multiple cluster head selected hence improves energy consumption.

### 2.2.2. Neighbourhood Based Clustering Algorithm:

Neighbourhood based clustering algorithm considers the information provided by the neighbouring nodes and on the basis of this information a cluster head is selected. Highest-Connectivity Cluster Algorithm (HCCA) (Kumarawadu, Dechene, Luccini, & Sauer, 2008) is based on selecting the node with highest neighbourhood nodes at the distance of 1-hop as cluster head. Another is Max-Min D-Cluster Algorithm (Amis, Prakash, Thai, Dung, & Huynh, 2000) which chooses the cluster head in a manner that none of its neighbouring nodes are at a distance of d-hop. The HCCA requires the strict clock synchronization while the Max-Min D-Cluster Algorithm has a benefit that it does not requires any clock synchronization.

Weighted Clustering Algorithm (Maniakchatterjee, K.das, & Turgut, 2002) is also a neighbourhood based clustering algorithm. It checks if there is any sensor node which has lost its connectivity with cluster head if so it reconfigures the network while keeping the balance of network. Grid-clustering ROUting Protocol (GROUP) (Yu, Wang, Zhang, & Zheng, 2006) has multiple sinks and the primary sink selects the cluster head on neighbourhood basis.

### 2.2.3. Probabilistic Clustering Algorithms:

Probabilistic Algorithms for clustering depends in the probability values of the nodes in WSN. LEACH is a probabilistic algorithm (Loscrì, Morabito, & Marano, 2005). Two-Level LEACH (TL-LEACH) is an extension of LEACH which has a new feature of two levels of cluster heads: primary level and secondary level. This protocol is feasible when the number of deployed nodes is very high.

Energy Efficient Clustering Scheme (EECS) (Ye, Li, Chen, & Wu, 2005) uses the dynamic, localized and non-repetitive competition based selection of cluster heads in WSN. Hybrid Energy Efficient Distributed Clustering (HEED) considers the energy of the sensor nodes and the cost of communication between any two nodes linked to same cluster head and chooses the cluster head in multi-hop networks (Younis & Fahmy, 2004).

### 2.2.4. Iterative Clustering Algorithm:

Iterative clustering protocols are SPAN, Algorithm for Cluster Establishment (ACE) and Distributed Clustering Algorithm (DCA). SPAN (Chen, Jamieson, HariBalakrishnan, & Morris, 2002) uses random cluster head selection but has a localized decision making based on the number of neighbour nodes allotted to cluster head and its energy level. ACE (Chan & Perrig, 2004) has two phases: spawning phase for new cluster and for existing cluster there is migration phase. DCA is a better iterative clustering algorithm as compare to SPAN and ACE. DCA protocol (Kumarawadu, Dechene, Luccini, & Sauer, 2008) uses the late announcement for willingness methodology for any sensor node before becoming the cluster head. Its advantage is that higher-weighted neighbour nodes have a chance to become cluster head.

### 2.3. Centralized Clustering Algorithm:

Centralized Clustering algorithms are also used in WSN. Following are some commonly used centralized clustering algorithm:

#### 2.3.1. LEACH (Low-Energy Adaptive Clustering Hierarchy) protocol:

LEACH was introduced by Heinzelman et al.,(Heinzelman, Chandrakasan, & Balakrishnan, 2000) which is used for the self-organized heuristic clustering. By self-organized it means that the network is divided into several clusters and the division of network is done on some criteria by the network as it organizes itself for better energy utilization. Each cluster of nodes has a Cluster Head node and many other nodes.

LEACH is the first hierarchical protocol of WSN based on data fusion. Cluster-based protocols like LEACH-C, LEACH-F, TEEN and PEGASIS are based on LEACH. Circular mode LEACH operation is a "Round" which consists of set-up phase and steady phase. In set-up phase, CH is generated randomly by selecting any number in a range between 0 and 1 in each sensor node. If the random number selected is smaller than threshold T (n), then the selected node is considered as CH. Mathematical equation for T (n) is as follows:

$$T(n) = \begin{cases} \dfrac{p}{1 - p[r \bmod \left(\dfrac{1}{p}\right)]} & n \in G \\ 0 & Other \end{cases}$$

Where, p is the percentage of CH in WSN, G is the cluster node set except CH of the last 1/p rounds while r is the current round number.

CH node broadcasts the message that it has been selected as CH to the entire WSN, all the nodes then decides to join which CH based on the strength of information received. Each node respond to the respective CH it selects. The data collected by the nodes is first transmitted to the CH and then the CH using the TDMA protocol sent the collected data to the sink node. Among the clusters, each cluster contends communication channel with CDMA protocol.

After the second phase (steady phase), the WSN enters the next round of the cycle again. The method of CH random selection reduces energy consumption thus increases the WSN life-span; data collection reduces the traffic efficiently (Xu & Gao, 2011).

For solving the disadvantage of formula in terms of T (n) in LEACH, an algorithm DCHS (Deterministic Cluster-Head Selection) was introduced as it adds the energy factors and thus enhanced the formula of T (n) for LEACH (Handy, Haase, & Timmermann, 2002). Limitation of DCHS algorithm is that if a node has very low energy and is selected as CH of the nodes then it will quickly die out which is not suitable for WSN. Nodes of WSN require a high power communication therefore expansion is limited. Nodes far from the sink node and yet communicating with each other at high power will result in shorter lifetime of WSN. The repetitive selection of CH will lead to the traffic and which results in energy consumption.

LEACH is classical protocol which is more energy efficient as compare to simple static network clustering algorithm and multi-hop routing protocols (Gang, Dongmei, & Yuanzhong, 2007).

For sensor nodes having long life-span a protocol has been proposed Energy Aware Routing Protocol (EAP). The main motivation for the proposed protocol was the issue of balancing the energy for sensor nodes according to applications requirement. EAP works better as it minimizes the energy consumption and load balancing for communication between the sensor nodes in return increases the life-span of the sensor nodes. A parameter for clustering was introduced for improved CH selection as it can effectively handle the heterogeneous energy capacities of WSN nodes. EAP clusters the sensor nodes into sets and form routing tree among various CHs for efficient energy consumption (Ming, Jiannong, Guihai, & Xiaomin, 2009).

Energy and Distance LEACH (EDL) was proposed to take residual energy and distance of sensor nodes into account as key parameters (Hou, Ren, & Zhang, 2009). EDL distributes CHs uniformly even in limited regions therefore the unnecessary energy consumption due to uneven distribution of CHs (short distance between two CHs) is reduced. EDL protocol is appropriate for heterogeneous as well as homogeneous networks.

#### 2.3.2. LEACH-C (Low-Energy Adaptive Clustering Hierarchy Centralized):

The set-up phase of the LEACH was improved as the cluster head random selection in each round has no great impact on performance. A central control algorithm was used to divide the WSN into clusters.

By distributing the cluster head in the WSN the network was centralized. This algorithm minimized the amount of energy for the nodes to transmit their collected data to the cluster head by minimizing the sum of squared distances of the nodes with their closest cluster head (Heinzelman, Chandrakasan, & Balakrishnan, 2002).

LEACH-C selects CH based on global information, which solves the problem of LEACH as it does not take global information into account (Xu & Gao, 2011). LEACH-C is centralized clustering algorithm which proposes transmission of data collected by nodes by location and energy levels to the base stations. Those nodes with high energy levels are considered as the cluster head (Xinhua & Sheng, 2010).

*2.3.3. LEACH-F(Low-Energy Adaptive Clustering Hierarchy Fixed):*

LEACH-F (Heinzelman W. , 2000) is similar to LEACH-C in cluster head selection method. But the clusters are fixed once they are formed whereas the cluster head position is taken by the different nodes within a single cluster.

*2.3.4. PEGASIS:*

PEGASIS (Power-Efficient Gathering in Sensor Information Systems) is based on the LEACH. It uses dynamic selection of CH. To evade energy consumption of the communication of frequent selection, all nodes in WSN are placed into structure chain (Lindsey & Raghavendra, 2002).

As the nodes are in straight chain therefore they know the location of neighbouring nodes. For sending and receiving of data nearest neighbours are calculated by the help of greedy algorithm. Data is transmitted in end to end form and then transmitted to sink node.

In figure 4 if node X is the CH then it broadcasts the message i.e. CH flag to all nodes, node V receives the CH flag and transmits its data to the node W. Node W will combine its data and node V data and transmit to CH. In the same manner, node Z will transmit its data to node Y and node Y will combine its own data along with Z node's data and forward it to CH i.e. node X. Node X will receive data from neighbouring nodes and it will also combine its own data with neighbours data and transmit it to the sink.

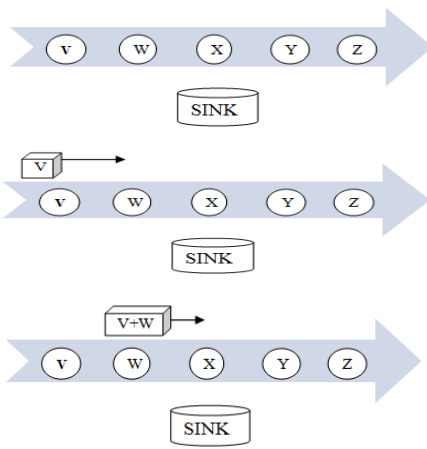

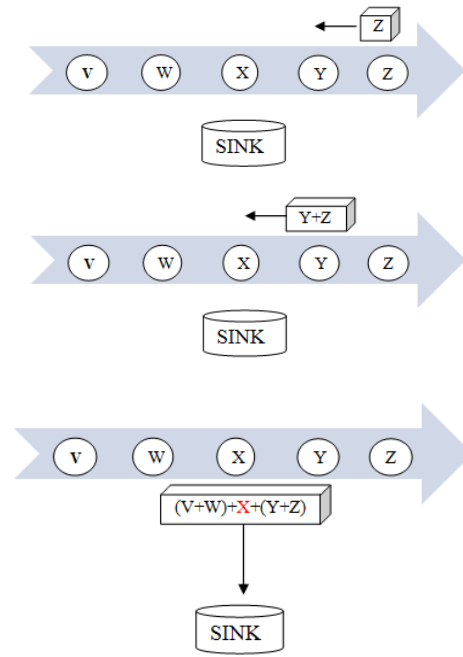

Figure 4: Data Collection in PEGASIS (along the chain)

Formation of chain structure and the CH within the chain is the crucial step. If the chain is large then all nodes should have the capability to communicate with the sink node thus resulting in transmission delay. As compared to LEACH, the PEGASIS reduces communication cost as well as the number of data transmission and communication volume through the chain of data aggregation.

An enhanced PEGASIS is introduced as Layered PEGASIS (Lindsey, Raghavendra, & Sivalingam, 2001). It improves the data transmission delay in long chain. There are two methods used for avoiding the data interference and node conflicts. First method is use of signal coding like CDMA. Second method is to allow a subset of nodes to transmit data at same time. Layered transmission tree is formed. The top level node is selected to communicate with each level (layer) for parallel data transmission and transmission delay reduction.

*2.3.5. SolarLeach Clustering Algorithm:*

The idea of cluster head has also been used in the SolarLeach clustering algorithm by T. Voigt et al. (Voigt, Dunkels, Alonso, Ritter, & Schiller, 2004).

*2.3.6. Base station Controlled Dynamic Clustering Protocol (BCDCP):*

Base station Controlled Dynamic Clustering Protocol (Murugunathan, Daniel, Bhasin, & Fapojuwo, 2005) depends on the base station for the selection of the cluster heads from a number of sensor nodes by using threshold for the energy level of the node for selection as sensing mode for simple non-cluster head nodes and cluster mode for those nodes which are selected as cluster nodes.

BCDCP enhances the durability of the network. It has two modes i.e. sensing mode and CH mode. Sensor nodes sense the assigned task and transmit it to the CH in sensing mode. From base station to CH all data gathering, data fusion and data transmission is done in CH mode. Those sensor nodes which have more residual energy than the average energy of the WSN are eligible for being CHs. Base Station formed the clusters by the help of iterative cluster splitting algorithm. BCDCP distributes the energy dissipation equally between all the sensor nodes. This even distribution improves network life-span and reduces average energy consumptions.

*2.3.7. Dynamic Minimum Spanning Tree Routing Protocol (DMSTRP):*

Dynamic Minimum Spanning Tree Routing Protocol (DMSTRP) (Huang, Xiaowei, & Jing, 2006)improves the BCDCP by using the Spanning tree concept for optimal decisions regarding the intra-clusters as well as inter-clusters. It is more useful for the large networks as compare to the LEACH-C and BCDCP because it always makes the optimal decisions considering the nodes within the cluster as well as the nodes outside it.

*2.3.8. HCR Algorithm:*

The nodes in the network are introduced in a round robin technique to the Cluster Head (Martin & Hussain, 2006). The only fixed positioned nodes in network are the sensor nodes (Martin & Hussain, 2006).

*2.3.9. Node Localization Using Genetic Algorithm:*

K-means method was used for clustering of nodes (Romoozi & Ebrahimpour-komleh, 2012). In K-means algorithm sufficiency was checked in terms of two factors which are: network coverage and the energy of transmission. The efficiency of the network localization is measured by the energy of transmission. The equation used is:

$$E = \sum_{j=1}^{k} \varepsilon_i \times E_{T_{jk}} + k \times E_R + E_{T_{hs}}$$

Where k is number is number of nodes in cluster, $\varepsilon_i$ is network traffic co-efficient, node's transmission energy is given by $E_{T_{jk}}$ while $E_R$ gives the energy of receiving packets by cluster head. The factor $E_{T_{hs}}$ gives the energy transmission from cluster head to sink. The node positioning was done by genetic algorithm. This increased the network's life-span and the network coverage was maintained in a better way.

*2.3.10. Genetic Algorithm Based Energy Efficient Clusters (GABEEC):*

For the optimization of the lifespan of the WSN by means of round by genetic algorithm the GABEEC was proposed (Selim Bayraklı, 2012). Static clustering with dynamic cluster head selection methodology is used. It has two steps: set-up and steady-state phase. The GABEEC method expands the lifetime of the network. Method starts with randomly selected nodes as CH of the network. Then by using genetic algorithm the location and the number of cluster heads are determined which are energy-efficient.

*2.3.11. TEEN*

TEEN (Threshold sensitive energy efficient sensor Network protocol) uses clustering algorithm resembling that of LEACH. TEEN is specifically designed for reactive wireless sensor networks (Manjeshwar & Agrawal, 2001). For cluster formation there are two main thresholds i.e. hard and soft. TEEN reduces the data transmission by the help of filtering method. Once the CH is selected, the two thresholds are broadcasted to all nodes along with the data transmission by TDMA. By setting the soft and hard threshold repeatedly the amount of data transmission is reduced. The problem in TEEN is that the threshold prevents some data transmission as it monitors the hot spots as well as unexpected events. TEEN is not a good choice for networks which require intervallic data reporting.

APTEEN (Adaptive Threshold sensitive Energy Efficient sensor Network protocol) is based on TEEN protocol. APTEEN (Agrawal & Manjeshwar, 2002) which can change related parameters issued by the CH according to the needs of users of WSN or type of use such as operation mode (TDMA); set of material attributes which users need; soft and hard threshold and the counting time (CT) representing time for complete data transmission of any node.

*2.3.12. DECSA (Distance-Energy Cluster Structure Algorithm):*

Distance-energy cluster structure algorithm is based on the clustering algorithm LEACH. DECSA takes distance as well as residual energy of nodes into consideration. Cluster head selection and cluster formation steps are improved as compared to classical LEACH clustering algorithm. Due to non-uniform distribution of the nodes within a WSN adverse effect on cluster head energy consumption have been observed. DECSA reduces these adverse effects on energy utilization of the cluster head. By avoiding direct communication of base sensor (BS) with cluster head (which is at distance from BS) objective of low energy consumption is achieved. DECSA algorithm effectively balances the energy consumption of WSN (Yong & Pei, 2012).

*2.3.13. ECHSSDA (Efficient Cluster Head Selection Scheme for Data Aggregation protocol):*

ECHSSDA has two phases: set-up phase and steady phase. For the improvement of the life-span of the network the better cluster head selection method was required (Maraiya, Kant, & Gupta, 2011). In ECHSSDA, after a round's completion the base-station monitors the energy of the current cluster head and the remaining nodes' average energy. If the cluster head energy is less than the average energy of the nodes then an associate cluster head will be selected.

This method has more energy-efficiency as compared to LEACH and LEACH-C. A systematic review about the LEACH protocols for clustering and routing is done by Kumar et al. (Tyagi & Kumar, 2013).

There is a requirement of security in LEACH protocol; this has motivated researchers to propose secure versions of LEACH protocol and make it safe from attacks of insider and outsider.

Review has been done in 2013 on the analysis of various secure LEACH clustering protocols in wireless sensor network (Masdari, Bazarchi, & Bidaki, 2013).

TABLE I
COMPARISON OF DIFFERENT CLUSTERING ALGORITHMS

| Algorithm | Researchers | Year | Function | Advantages | Limitations |
|---|---|---|---|---|---|
| LEACH (Heinzelman, Chandrakasan, & Balakrishnan, 2000) | Heinzelman, W.B. | 2000 | WSN self-organizing for efficient energy consumption | More energy efficient than static network clustering algorithm and multi-hop routing protocols. | Location of nodes not considered. Node with low energy if selected as CH will die quickly. |
| LEACH-C (Heinzelman, Chandrakasan, & Balakrishnan, 2002) | Heinzelman, W.B. | 2002 | Minimize energy of nodes by minimizing sum of squared distances of the nodes with closest CH | LEACH Set-up phase improved by use of central control algorithm to divide the WSN into clusters | Disproportion in energy consumption rates between various nodes in the network not considered. |
| LEACH-F (Heinzelman W., 2000) | Heinzelman W. | 2000 | CH selection similar to LEACH-C. CH changes within cluster while clusters are fixed. | LEACH-F cluster will not be constructed repeatedly reducing cost of cluster formation | LEACH-F cannot handle dynamic node joins, failures and movement. Signal interference between clusters is increased. |
| TEEN (Manjeshwar & Agrawal, 2009) | Manjeshwar A. et al. | 2002 | Protocol for Enhanced Efficiency in WSN | By setting soft and hard threshold repeatedly amount of data transmission is reduced. | Threshold hinders data transmission as it monitors unexpected events. TEEN is not used in networks requiring periodic data reporting |
| APTEEN (Manjeshwa & Agrawal, 2002) | Manjeshwar A. and Agrawal D.P. | 2002 | Hybrid protocol for efficient routing and comprehensive information retrieval | Parameters like operation mode, soft and hard threshold, and counting time (CT) issued by CH are adjustable | It changes issued parameters for user requirement but if chnaged repeatedly it will be an overhead. |
| PEGASIS (Lindsey & Raghavendra, 2003) | Lindsey S. et al. | 2003 | PEGASIS has improved network lifetime than LEACH but not suitable for large WSN | Nodes are in chain structure so they may operate in low power mode. Reduced number of data transmission. | If the chain is long then transmission delay increased. Needs to know location of other nodes. |
| HEED (Younis & Fahmy, 2004) | Younis O. et al. | 2004 | HEED periodically selects CH based on node's energy and node degree | Energy of the sensor nodes and communication cost of any two nodes having same CH is considered. It has O(1) iterations with minimum message overhead. | Requires upper bound on intercluster and intracluster transmission ranges. Node density range is also pre-defined |
| EEPSC (Amir Sepasi Zahmati, 2007) | Amir Sepasi Zahmati et al. | 2007 | Energy-Efficient Protocol with Static Clustering | Uses dynamic, localized and non-repetitive competition based selection of increasing WSN lifetime. | High energy nodes in comparison to normal nodes are not considered in cluster if Base Station is at distance. |
| MRPUC (Bencan Gong, 2008) | Bencan Gong et al. | 2008 | Unequal clustering and routing of multi-hop for better lifetime of WSN, additional overhead for multi-hop. | Selects CH which has more residual energy closer to BS. Distance and residual energy checked when node joins cluster. | Relay nodes have maximum residual energy while having minimum energy utilization for forwarding packets which is inefficient method of relay node selection. |
| ECHSSDA (Maraiya, Kant, & Gupta, 2011) | Maraiya, Kant and Gupta | 2011 | After each round completion BS compares the energy of the current CH with the remaining nodes' average energy | ECHSSDA is more energy-efficiency as compare to LEACH and LEACH-C. | If CH energy is less than average energy of the nodes then an associate CH will be selected which is an overhead. |
| GABEEC (Kuila & Jana, 2012) | Pratyay Kuila and Prasanta Jana | 2012 | Random selection of CH. Using GA for location and node degree of CH | Use of genetic algorithm (GA) for location and the number of cluster heads | Random selection of nodes as cluster head of the network. |

| | | | | | which are energy-efficient | |
|---|---|---|---|---|---|---|
| DECSA (Yong & Pei, 2012) | Yong and Pei | 2012 | Improved Cluster head selection and cluster formation. | It takes distance as well as residual energy of nodes into consideration. | Not found | |

### 2.4. Load-Balanced Clustering Algorithm:

A load Balanced clustering algorithm was proposed which has improved the WSN in terms of load balancing; energy consumption and execution time (Gaurav & Younis, 2003).

#### 2.4.1. Energy Efficient Balanced Clustering(EELBC) Algorithm:

Scalability is a major concern regarding the Wireless Sensor Network. EELBC algorithm not only improves the lifespan of the network but also improves the scalability of the network (Kuila & Jana, 2012). It is based on a min-heap clustering algorithm. Using the cluster heads a min-heap is build based on the nodes which are allotted to that particular cluster head.

#### 2.4.2. Bipartite Graph for Load Balanced Clustering:

The use of bipartite Graph of sensor nodes and gateways for selection of CH for a sensor node is also a method for load balanced clustering. Its time complexity is $O(mn^2)$ but it is not very effective in case of large WSN as the execution time will be high (Chor Ping Low, 2008).

#### 2.4.3. Improved Load Balanced Clustering Algorithm:

Another algorithm was proposed which took care of the issues like load balancing along with energy consumption (Pratyay & Prasanta, 2012). It is more efficient in terms of dead sensor nodes and time complexity which is: $O(n \log m)$

Some other research works on the load balanced clustering algorithm are done by Wei Li, (Li W., 2009) Xiang Min et al.,(Xiang Min, 2010) and Zhixin Liu et al. (Zhixin & et, 2011). Another novel work is on new cluster-based protocol named ROL (Route Optimisation and Load-balancing). It uses many Quality of Service (QoS) metrics to meet the application requirements of wireless sensor network (Hammoudeh & Newman, 2013).

### 3. ENERGY EFFICIENT ROUTING IN WSN

Sensor nodes in WSN are mostly smaller in size but are energy constrained.

WSN are future generation of sensor networks having numerous applications. Single routing protocol cannot fulfil the need of different applications; therefore many routing protocols have been introduced. According to network topology, routing protocols are categorized into flat and hierarchical routing protocol (Elson & Estrin, 2004). Data collection method may also improve energy efficiency to extend the lifetime of the network. Better routing techniques have been able to provide faster communication with minimum energy consumption within a WSN.

#### 3.1. Unit Disk Graph (UDG):

Unit Disk Graph (UDG) is a graph for representing the nodes in WSN. All the sensor nodes have same energy levels and can be devised as a unit disk graph (Seetharam, Bhattacharyya, & Chakrabarti, 2009).

If the Euclidean distance between two nodes is at maximum one then the edge will be present between these two nodes (Li X.-Y., 2003). A Wireless Sensor Network can never have a fixed infrastructure. As it cannot be predefined therefore the communication of nodes may be of single or multi-hop as they have shared medium (Thai, Wang, Zhu, & Zhu., 2007).

#### 3.2. Minimum Connected Dominating Set (MCDS):

Even with no predefined infrastructure a virtual outline structure can be created by Minimum Connected Dominating Set (MCDS) which will minimize the communication overhead, improve the connection efficiency while reducing the energy consumption and extending network's lifetime (Rai, Verma, & Tapaswi, 2009).

#### 3.3. Maximal Independent Set (MIS):

Another method is Maximal Independent Set (MIS) is used when any node not in dominating set has a neighbor in dominating set. Firstly the dominating set was to be formed and afterwards its connections in form of vertices are added (Wua, Dub, Jia, Li, & Huang, 2006).

#### 3.4. One round MCDS Algorithm:

Clustering Technique is useful for the data collection from nodes to the cluster head of each cluster. It will minimize the data-redundancy issue as well as the communication overhead. When there is no neighborhood information then the most suitable method is to solve the minimum dominating set problem as proposed in One round MCDS algorithm (Gao, Guibas, Hershberger, Zhang, & Zhu, 2001). Its time complexity is (O√n).

#### 3.5. Fast Distributed MIS:

A better method was required which resulted in the research work of a fast distributed MIS which works in round of phases (Luby, 1985). Its time complexity is O (log n). But the disadvantage is that the distances are not known which increases the computation.

#### 3.6. Energy Efficient Sensor Network Framework (EESNF):

The distance knowledge is known in the Energy Distributed Maximal Independent set with Known Distances (EDMISKD). It is an Energy Efficient Sensor Network Framework (EESNF) which will enable certain energy-efficient routing levels to be achieved in WSN. Three algorithms have been proposed: Efficient distributed maximal independent set, Energy aware data routing technique (EADRT) and Balanced shortest path tree for data gathering (BSPTDG) (Das, Barman, & Sinha, 2012).

Some recent work is being done on special case for specific type of clustering a particular routing strategy maybe proposed. A novel routing protocol is designed specifically for multi-hop clustered wireless sensor networks which is named as flow-balanced routing (FBR) protocol. It is composed of four algorithms. These are network clustering algorithm, multi-hop backbone construction algorithm, flow-balanced transmission algorithm, and rerouting algorithm. The clustering algorithm classifies several sensors into single cluster on the overlapping degrees of these sensors. Algorithm for backbone construction creates new multi-level backbone by the help of sink and cluster-heads. Algorithm for flow-balanced routing assigns the transferred data to multiple paths equalizing the power consumption of all the sensors in wireless sensor network. This is used to attain power efficiency as well as coverage preservation (Tao, Zhang, & Ji, 2013).

4. ENERGY EFFICIENT MAC PROTOCOLS FOR WSN

Energy Efficient MAC protocols have been proposed for wireless sensor networks (WSNs). Nodes have limited power hence at the MAC level energy efficiency is of great importance. The nodes communicate with BS and other nodes via radios while the BS is a gateway between the nodes of WSN to the application for exchanging data. Energy efficiency is essential requirement which affects the MAC layer protocol for WSN (Society, 1997). Duty Cycling is a primary technique for energy constrained WSN for achieving low energy operation (Anastasi, Conti, Di, & Passarella, 2009).

MAC layer should be able to minimize the radio energy costs in nodes so that the problem of low power radio is solved. As MAC layer can directly control radio energy therefore it is more effective. MAC protocol is used in WSN as it can maximize the sleep duration of nodes; minimize the idle listening, sensing, overhearing among nodes and removing hidden terminal issues or packet collisions (Stankovic & He, 2012). A MAC protocol for WSN has the attributes shown in figure 4.

Existing MAC protocols can be categorized as asynchronous and synchronous MAC protocols. Duty cycling MAC protocols are also called synchronous energy efficient MAC protocols.

Time synchronization is required therefore Time slot are formed.

Some popular MAC protocols are given in precise detail.

*4.1. Sensor MAC (SMAC):*

Sensor MAC (SMAC) was the first synchronous energy efficient MAC protocol designed for WSN. Energy saving relies on the duty cycle. It uses three techniques for energy consumption's reduction:

a) Periodic Sleep:

Locally managed synchronizations and periodic sleep scheduling is done in this technique. Periodic sleep results in high latency for multi-hop algorithm for routing.

b) Virtual Clustering:

Virtual Cluster is formed by the neighbouring cells to setup a similar sleep schedule.

c) Adaptive Listening:

Adaptive listening is a technique proposed for sleep delay in return for latency of the WSN.

In SMAC, neighboring nodes wake-up simultaneously and listen for channel. Disadvantage is the collision avoidance back-off and the RTS/CTS exchange for the collision free transmission of data from neighbors.

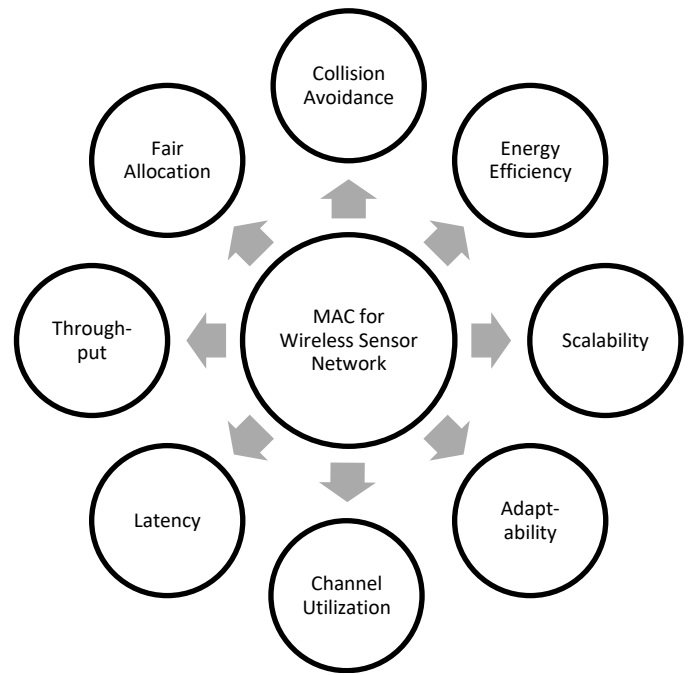

Figure 5: Wireless Sensor Network MAC Attributes (Roy & Sarma, 2012).

Another disadvantage is that it fixes the duty cycle so it is not an optimal solution and energy wastage for idle-listening when the data-rate are low (Ye, Heidemann, & Estrin, June 2002).

### 4.2. Timeout MAC (TMAC):

Timeout MAC (TMAC) lessens the uptime of SMAC by the help of a timer (Dam & Langendoen, November 2003). It has adaptive duty-cycle. TMAC thus reduces the uptime when channel is in idle state. In TMAC nodes transmit messages in bursts of varying length and sleep mode between these bursts. It also uses the RTS/CTS/ACK scheme. TMAC improves SMAC as it reduces the awake-duration when the channel is idle. Disadvantage is that it has an early sleeping issue which occurs when a node goes to sleep mode when a neighbor is still sending messages to that node. This problem arises due to the asymmetric communication. By use of Future-Request-To-Send (FRTS) this problem is solved.

### 4.3. Scheduled Channel Polling MAC (SCPMAC):

Scheduled Channel Polling MAC (SCPMAC) is used to minimize the preamble by the help of combination of scheduling techniques along with the preamble sampling. It finds optimal parameters for periodic traffic in WSN but it does not reduce the overhearing energy loss. The result of which is more delay and contention. Variable traffic is adapted by SCPMAC. In case of heavy traffic duty cycle are increased. It is similar to SMAC and TMAC. Advantage is that it adds extra and high-frequency polling slots to the sensor nodes of WSN on the path (Ye, Silva, & Heidermann, 2006).

### 4.4. Adaptive Energy Efficient MAC Protocol (AEEMAC):

Adaptive Energy Efficient MAC Protocol (AEEMAC) uses a duty cycling to reduce energy consumption as it avoids idle listening for channel (Roy & Sarma, 16-18 Dec. 2011). In AEEMAC protocol three optimization techniques are introduced: adaptive sleeping, two combined transmission schemes and reusing of channel scheme. It has better performance as compare to SMAC. It improves the end-to-end delay, throughput and packet delivery. It reduces the energy consumption in WSN.

Few other MAC protocols for energy-efficiency in WSN are DMAC,(Lu, Krishnamachari, & Raghavendr, 2004) TDMA MAC protocols, (Ray, Demirkol, & Heinzelman, 2010) Interleaved Spatial Temporal (IST) Scheduling,(Barnawi & Hafez, 2008) Data Aggregation TDMA Protocol (DATP),(D´ıaz-Anad´on & Leung, 2011) Hybrid TDMA MAC protocols,(Sitanayah, Sreenan, & Brown, 2010) Z-MAC,(Rhee, Warrier, Aia, Min, & Sichitiu, 2008) B-MAC (Polastre, Hill, & Culler, 2004) and Adaptive TDMA MAC protocol (Barnawi A. Y., 2012).

A comprehensive comparison of the ATMA protocol with S-MAC, T-MAC, ADV-MAC and TDMA-based protocol (TRAMA) as well as the Z-MAC and IEEE 802.15.4 which are both hybrid protocols through extensive simulations and qualitative analysis is given by Ray et al. (Ray, Demirkol, & Heinzelman, 2013).

They showed that ATMA outperforms all the existing protocols in terms of energy consumption and latency for several simulation scenarios which were studied.

## 5. ENERGY EFFICIENT DATA AGGREGATION ALGORITHMS FOR WSN

Communication overhead is an important issue for the WSN. For the reduction of the communication overhead a technique used is known as Data Aggregation. It helps in expanding the lifetime of the network. WSN are used in pervasive computing due to its flexibility and less-cost. Human interaction is involved which requires privacy of the data aggregation. Data aggregation should have less communication overhead. Accuracy of data aggregation should be high.

### 5.1. Slice-Mix-AggRegaTe (SMART):

The issues such as the energy constraint, insecurity and lack of protection are needed to be minimized. Thus a scheme was proposed which is known as Slice-Mix-AggRegaTe (SMART). It ensures the data privacy by data slicing technique.

### 5.2. Energy Efficient and High- Accuracy secure data aggregation (EEHA):

Another technique is Energy Efficient and High- Accuracy secure data aggregation (EEHA). In this technique the data slicing and mixing are used but only on the leaf nodes. All the data of the leaf nodes is transmitted to the aggregator nodes which are intermediate nodes (Cam, Ozdemir, Nair, Muthuavinashiappan, & Sanli, 2006).

### 5.3. Witness based SMART (WSMART):

A variation of SMART technique is introduced which is called Witness based SMART (WSMART). It removes the problem of high communication overhead and data-packets collision. Its advantage is that it validated the result of the data aggregator.

WSMART gives data protection with less overhead than the SMART and EEHA scheme as it has low energy consumption and minimum bandwidth utilization. It also has more accuracy level than both SMART and EEHA (Vinoth & Kumar, 2012).

### 5.4. Secure Information Aggregation (SIA) in Sensor Networks:

It is one of the first researches done for secure information aggregation in sensor networks (Przydatek, Song, & Perrig, 2003). This technique was used to overcome the intruder aggregator and sensor nodes. Basic framework was that there were some nodes called aggregators which aggregate the data requested by the query. It minimizes the communication overhead. Disadvantage is that the data of all the nodes has to be transmitted to the base-station for aggregation therefore increasing communication overhead.

### 5.5. A Witness Based Approach for Data Fusion Assurance in WSN:

For the information collection nodes are spread out over the network coverage area. There are some witness nodes which validate the aggregation results as well as calculates the MAC value of the result and forwards it to aggregator node which forwards it further to the base station (Du, Deng, Han, & Varshney, 2003).

### 5.6. A Secure hop by hop data aggregation (SDAP) protocol for Sensor Networks:

It is based on divide and conquer rule. SDAP utilizes probabilistic technique for grouping of nodes in a tree form. Hop-by-hop aggregation is done in each node group to create an aggregate group. Disadvantage in this method is the attestation process when it is proving the validation of its group aggregate (Yang, Wang, Zhu, & Cao, 2006).

### 5.7. Privacy Preserving Data Aggregation (PDA):

SMART technique is a technique for PDA. It ensures the data privacy by slicing and assembling techniques. Every node slices its data into smaller pieces for privacy measures. The intermediate node when receive this sliced data it calculates aggregate and transmit it to sink (He, Liu, Nguyen, Nahrstedt, & Abdelzaher, 2007).

### 5.8. Efficient Aggregation of encrypted data in Wireless Sensor Networks:

WSNs are formed by small devices having limited computation and low-energy. Data transmission requires large energy consumption. A more secure and simple technique is to use homomorphic stream cipher (Castelluccia, Mykletun, & Tsudik, 2005). It allows efficient data aggregation. Its disadvantage is that it increases the overhead by transmitting the identities of node along with the aggregate to the sink.

Some other algorithms for data aggregation are Enhanced Directed Diffusion (EDD),(Li, Lin, & Li, 2010) Power Efficient GAthering in Sensor Information Systems (PEGASIS),(Huang, Shieh, & Tygar, 2009) Energy Aware Distributed Aggregation Tree (EADAT), Cooperative data aggregation (CDA),(Tan & Korpeoqlu, 2003) DMCT, Isoline data aggregation scheme(Solis & Obraczka, 2009) and improved Isoline data aggregation scheme (Guocan & Guowei, 2011).

A recent research work is done on the energy-efficient and balanced cluster based data aggregation. EEBCDA is an algorithm proposed for the division of network into rectangular unequal grids. It has enhanced energy efficiency and expanded the lifetime of network. It has also solved unbalanced energy dissipation problem in WSN (Yue, Zhang, Xiao, Tang, & Tang, 2012).

A novel research is done on an operator placement problem in Wireless Sensor Networks (WSN). This operator placement assignment is defined on a particular node of the network each (query) operator will be hosted and executed. Therefore operator placement problem has to be addressed as it has crucial role in the terms of query optimization issues in wireless sensor network. It is very relevant for in-network query. Any query routing tree has to be decomposed into three sub-components that have to be processed at the time of query. These three components are operator tree, operator placement assignment method and routing scheme (Chatzimilioudis, Cuzzocrea, Gunopulos, & Mamoulis, 2013).

In 2013, a novel work was done in real-time wireless sensor networks for energy efficient ways by exploiting data redundancy. They developed scheduling algorithms for energy efficiency in Real time wireless sensor networks (WSNs) by controlling the energy-delay tradeoff (Fateh & Manimaran, 2013).

There are some other fields such as querying or tracking approach which vary from one wireless sensor network to another. A survey related to Querying and tracking services is done by Zuhal Can and Murat Demirbas (Can & Demirbas, 2013). They categorized querying techniques into geometrical based, tree-based, hash based and hierarchical cluster-based. Other aspects like fault-tolerance, distance-sensitivity, scalability and energy efficiency are also taken into account.

## 6. CONCLUSIONS

The energy efficient techniques are widely used in all the networks especially in Wireless Sensor Network as the nodes have low energy. The nodes have limited batteries and the lifetime of the network depends on the nodes 'energy. In this survey paper, a comprehensive review is proffered on the various energy efficient methods for the enhancement of the lifetime of WSN. In WSN, energy efficient clustering methods are used which are mainly categorized as centralized clustering techniques and distributed clustering techniques. Comparison of different clustering methods in terms of functionality, advantages and limitation is done in this research. Energy efficient routing in WSN has been an intensive research area for the past decade. Several routing techniques for WSN are proposed which have been able to provide faster communication with minimum energy consumption. Brief analysis of these routing techniques is presented. WSNs nodes have limited power hence at the MAC level energy efficiency is of immense importance. The improvement in the lifespan of wireless sensor network by using different energy efficient MAC protocols described.

Data Aggregation is a method used for reduction of communication overhead. This paper briefly discusses the techniques introduced for data aggregation and compares the various algorithms introduced for energy efficient WSN.